\documentclass[lettersize,journal]{IEEEtran}
\usepackage{amsmath,amsfonts, amsthm}
\usepackage{algorithmic}
\usepackage{algorithm}
\usepackage{array}
\usepackage[caption=false,font=normalsize,labelfont=sf,textfont=sf]{subfig}
\usepackage{textcomp}
\usepackage{stfloats}
\usepackage{url}
\usepackage{verbatim}
\usepackage{graphicx}
\usepackage{cite}
\newcommand{\me}{\mathrm{e}}

\newtheorem{innercustomthm}{Theorem}
\newenvironment{customthm}[1]
  {\renewcommand\theinnercustomthm{#1}\innercustomthm}
  {\endinnercustomthm}
  
\begin{document}

\title{Reduced Softmax Unit for Deep Neural Network Accelerators}

\author{Raghuram S,~\IEEEmembership{Member,~IEEE,}
\thanks{Raghuram S is an Associate Professor in the Department of Electronics \& Communication Engineering and the Co-ordinator, Center for Cyber-Physical Systems, M S Ramaiah Institute of Technology, Bengaluru 560054, India (e-mail: raghuram@msrit.edu).}}



\maketitle

\begin{abstract}
The Softmax activation layer is a very popular Deep Neural Network (DNN) component when dealing with multi-class prediction problems. However, in DNN accelerator implementations it creates additional complexities due to the need for computation of the exponential for each of its inputs. In this brief we propose a simplified version of the activation unit for accelerators, where only a comparator unit produces the classification result, by choosing the maximum among its inputs. Due to the nature of the activation function, we show that this result is always identical to the classification produced by the Softmax layer. 
\end{abstract}

\begin{IEEEkeywords}
Deep Neural Networks, Accelerators, Softmax Activation
\end{IEEEkeywords}

\section{Introduction}
\begin{figure}[!t]
\centering
\includegraphics[width=0.5\textwidth]{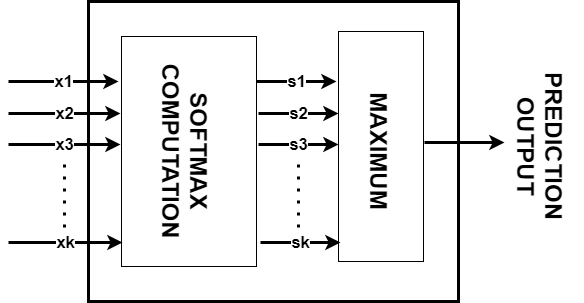}
\caption{Softmax Activation Layer}
\label{smax}
\end{figure}

AI models have greatly improved their classification accuracy primarily due to the power and flexibility of DNNs. While original efforts focused on image related classification problems, their application domain has extended to various diverse areas such as healthcare assistive technologies, natural language processing, reinforcement learning, etc. To increase the penetration of this revolutionary technology, hardware accelerators are being developed to improve the functionality of existing services and also to realize paradigms such as edge computing. Among the layers in a DNN, the Softmax Activation layer is the last or output layer and is always used when there are more than two classes among the inputs \cite{dlbook}. This result is typically a single integer value, representing the class of the input sampled. This layer is illustrated in Fig. \ref{smax}. 

An input sample, for example an image, is given to the input layer of the DNN, after which it passes through a series of layers as determined by the DNN architecture, before reaching the penultimate layer in the form of $x_i$ (Fig. \ref{smax}).  
If the classification problem is known to have $k$ possible classes, there will be $k$ values of $x_i$ which form the input to the softmax layer.
The softmax computation block computes the function defined in \eqref{smax_eqn}, and produces $k$ activations $s(x_i)$: each value represents the probability that the input sample belongs to class $i$. Hence, the prediction result of the DNN is the particular integer class $i$ which has the maximum value of $s(x_i)$ among all the $k$ activations. While this activation term poses no problems for software implementations, hardware accelerators require additional efforts to realize the exponential terms present in \eqref{smax_eqn}.   

\begin{equation}
\label{smax_eqn}
s(x_j) = \cfrac{\me^{x_j}}{\me^{x_1} + \me^{x_2} + \dots + \me^{x_k}}
\end{equation}

\section{Softmax Layer Implementations}
Hardware implementations depend on the simplification of the form of the activation function and then applying techniques such as lookup tables or approximations to estimate the exponential terms.

The authors in \cite{mocast} begin by applying a logarithmic operator to the activation function. This is acceptable, since the maximum value of the function is required, and applying the logarithm does not change the position where this maxima occurs. The logarithmic term is further simplified to obtain an approximation, where the exponential required to be calculated is bounded above by unity. It is argued that this is beneficial since the size of the LUT required is reduced. 

\cite{tran2020} modifies the base to 2 from $e$ in order to utilize the simplified operations in digital implementations. This can be achieved with the transformation shown in \eqref{2_e}, after which the exponent is expressed as a sum of an integer and a fraction. The integer is easily expressed as an appropriate number of shifts and the remaining $2^{v_i} : v_i < 1$ is realized with an LUT. An algorithm is presented to improve the quality of the approximation, dependent on a precision parameter P. A similar approach is followed in \cite{nature}, where a pseudo Softmax function is proposed by changing the base to 2 from $\me$.

\begin{equation}
    \me^{y_i} = 2^{y_i\cdot \log_2\me }
    \label{2_e}
\end{equation}

The authors in \cite{vdat} have used the CORDIC algorithm to calculate the exponential. In this work, an inverse softmax is also investigated, a function of the form \eqref{inv_smax}, which is nothing but the reciprocal function of \eqref{smax}. This modifies the predicted class to the one which produces a minimum value of $s'$. The advantage in this approach is that it avoids the requirement for the division operation.

\begin{equation}
\label{inv_smax}
s'(x_j) = 1 + \sum\limits_{\substack{i=0 \\ i\neq j}}^{N}\me^{x_i - x_j}
\end{equation}

\section{Reduced Softmax Activation Layer}
In DNN accelerators, due to the nature of the computations required, a von-Neumann type architecture is not practical. Hence, special techniques are devised to improve the efficiency of the inference process: accelerators typically implement optimizations at the algorithm, architecture, or circuit level \cite{dnn_acc}. For example, the authors in \cite{eyeriss} have designed an on-chip network to improve the utilization of the processing elements in the accelerator. In this section, we propose an algorithm level modification for the softmax activation layer, that results in a reduced circuit implementation. 

Each output $s(x_i)$ of the softmax activation function represents a discrete probability over the classes \cite{dlbook}. During the training process, these individual values are essential for the backpropagation algorithm loss computation, for example a cross entropy loss function as in \eqref{ce_func}, with target labels $t_i$.  

\begin{equation}
    E = - \sum\limits_{i}^{nclass}t_i\log s_(x_i)
    \label{ce_func}
\end{equation}

Accelerators that also include DNN training functionality are very few, compared to the accelerators that focus on inference \cite{rev2020}. This is primarily due to two reasons: the computational power required for the training process (done in specific hardware systems called as Graphics Processing Units or GPUs) is a few orders of magnitude higher than the computational power required for inference. More importantly, training is not done as frequently as inference. For example, a DNN model might be trained/re-trained after collecting data over a period of one year. However, inference might happen thousands of times during a day, for example, in an edge computing application. Hence, it is more appropriate for DNN accelerators to realize the inference functionality, as compared to the training capability. 

Inference engines are typically used as an embedded system in phones or cameras, or in stand-alone edge computing systems such as a face recognition system. The accelerator is responsible for efficiently obtaining the result of the classification task, as required in the application and determined by the architecture of the DNN. For these accelerators only the predicted class is required, i.e., the maximum of the $s(x_i)$ is required but the actual values are not necessary, as there is no requirement for loss computation to back-propagate the errors. 

\begin{equation}
    s(x_i) \propto \me^{x_i}
    \label{smax_eqn_red}
\end{equation}

When comparing the $s(x_i)$ values, the denominator in \eqref{smax_eqn} can be ignored, since it is a positive term and scales down all the $s(x_i)$ values uniformly. Hence, the maximum class is just the class for which $\me^{x_i}$ is largest. The search for the maximum candidate is further simplified by the fact that the exponential function is monotonically increasing, mapping inputs in $(-\infty, \infty)$ to $(0, \infty)$. 

\begin{customthm}{1}\label{xy_exy}

For any two real numbers $x$ and $y$, $x > y \longrightarrow s(x) > s(y)$ . 

\end{customthm}

\begin{proof}
Since $x > y$, we can write $t = x - y > 0$. Expanding $\me^t$ using Taylor's series expansion for exponentials:
\begin{equation*}
\me^t = 1 + \cfrac{t}{1!} + \cfrac{t^2}{2!} + \cfrac{t^3}{3!} + \cdots 
\end{equation*}

Since $t > 0$, $\me^t > 1$. Rewriting $t$ as $x-y$ we get: 

\begin{equation*}
   \me^{x-y} > 1 \longrightarrow \cfrac{\me^x}{\me^y} > 1 \longrightarrow \me^x > \me^y 
\end{equation*}
From \eqref{smax_eqn_red}, we get the relation
\begin{equation*}
    x > y \longrightarrow \me^x > \me^y \longrightarrow s(x) > s(y)
\end{equation*}

\end{proof}

\begin{figure}
\includegraphics[width = 0.45\textwidth]{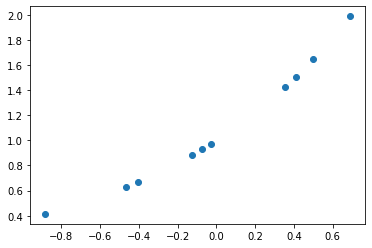}%
\begin{picture}(0,0)
\put(-200,80){\includegraphics[height=2cm]{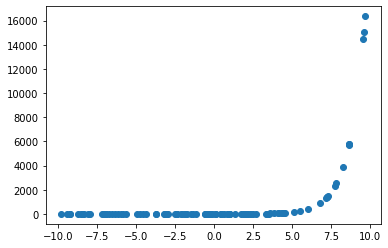}}
\end{picture}
\caption{Exponential values for a uniform random set of 10 inputs in [-1, 1] and  larger set from [-10, 10] (inset)}
\label{exp_vals}
\end{figure}

\begin{figure}
\includegraphics[width = 0.45\textwidth]{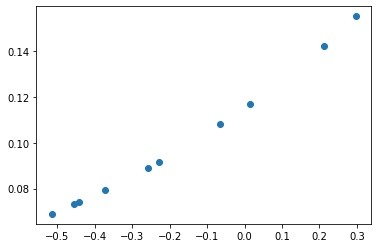}%
\begin{picture}(0,0)
\put(-200,80){\includegraphics[height=2cm]{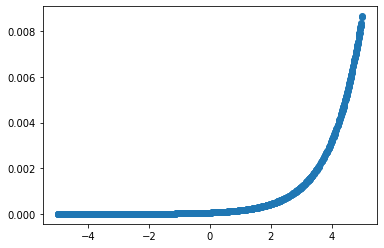}}
\end{picture}
\caption{Softmax values for a uniform random set of 10 inputs in [-1, 1] and  larger set from [-5, 5] (inset)}
\label{smax_vals}
\end{figure}

\begin{figure}[!t]
\centering
\includegraphics[width=0.45\textwidth]{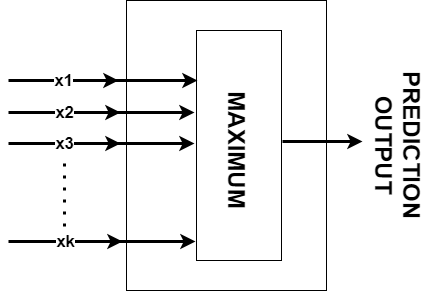}
\caption{Reduced Softmax Activation Layer}
\label{smax_red}
\end{figure}

From Thm. \ref{xy_exy}, for any two inputs $x_i$ and $x_j$, choosing the maximum among these two is equivalent to choosing the maximum among the softmax of the values. 
This is illustrated in Figures \ref{exp_vals} and \ref{smax_vals}, where uniformly sampled random inputs are shown along with their exponential and softmax values respectively. The monotonically increasing nature of the exponential, and hence the softmax, is clearly seen in the figures. Table \ref{smax_tab} shows three separate sets of inputs generated uniformly from the intervals [-100,0], [0, 100] and [-1, 1] along with the input, exponential, and softmax values: in each of the cases, we see that the input with maximum value corresponds to the highest probability as predicted by $s(x_i)$. At the circuit level, this translates to replacing the softmax activation with just a comparator, as shown in Fig. \ref{smax_red}.

\begin{table}[!t]
\caption{Softmax Output Samples}
\centering
\resizebox{\columnwidth}{!}{%
\begin{tabular}{|c|c|c||c|c|c||c|c|c|}
\hline
\multicolumn{3}{|c||}{All Negative (-100, 0)}& \multicolumn{3}{|c||}{All Positive (0, 100)}& \multicolumn{3}{|c|}{Random Inputs (-1, 1)}\\
\hline
Input&$\me^{x_i}$&$s(x_i)$&Input&$\me^{x_i}$&$s(x_i)$&Input&$\me^{x_i}$&$s(x_i)$\\
\hline

-67.98&2.98e-30&1.51e-25&62.31&1.16e27&3.80e-15&-0.95&0.38&0.03\\
-33.07&4.33e-15&2.19e-10&87.20&7.44e37&2.44e-04&-0.83&0.43&0.03\\
-76.26&7.54e-34&3.81e-29&10.66&4.27e04&1.41e-37&-0.69&0.49&0.04\\
-92.96&4.22e-41&2.13e-36&83.53&1.90e36&6.24e-06&0.58&1.79&0.16\\
-90.64&4.30e-40&2.17e-35&45.06&3.74e19&1.22e-22&-0.55&0.57&0.05\\
\bf{-10.83}&\bf{1.96e-05}&\bf{9.95e-01}& 73.87&1.20e32&3.95e-10 &0.16&1.18&0.10\\
-16.15&9.67e-08&4.89e-03&49.77&4.14e21&1.35e-20&0.23&1.26&0.11\\
-89.70&1.09e-39&5.55e-35&66.38&6.89e28&2.22e-13&\bf{0.91}&\bf{2.49}&\bf{0.22}\\
-36.38&1.57e-16&7.97e-12&23.36&1.40e10&4.61e-32&0.07&1.07&0.09\\
-60.84&3.75e-27& 1.89e-22&\bf{95.52}&\bf{3.05e41}&\bf{9.97e-01}&0.18&1.20&0.11\\

\hline
\end{tabular}} 
\label{smax_tab} 
\end{table}

\section{Discussion}
In this brief, we propose to replace the Softmax activation unit with only the comparator. 
In a majority of the accelerators, the focus is on quick inference, for example in edge applications.
In such systems, the DNN model is essentially static as there are no weight updates via the backpropagation algorithm. In these cases the probability of the classifications is not required, only the class predictions. The predicted class is chosen as the one which has highest probability, as computed by the softmax layer. We have shown that this layer's computation can be simplified by choosing the maximum of its inputs, as the softmax function preserves the ordering of the inputs before and after the activation is applied. This implies that the identical result is available, by just the ordering of the inputs and choosing the maximum among them as the predicted class, thereby obviating the need for the softmax function computation. This will greatly reduce the size of the unit in accelerators for applications such as object detection, where the output stage may be a 1000 class softmax layer. 

\end{document}